\documentclass{optica-article}

\journal{opticajournal} % for journals or Optica Open

\articletype{Research Article}

\usepackage{lineno}
\usepackage{here} % adds [H] option for figures to stay

\usepackage{amsmath,amsfonts}%

\usepackage{amsthm}%
\usepackage{physics}
\usepackage{manyfoot}%
\usepackage{mathrsfs}%
\usepackage{xcolor}%

\usepackage[normalem]{ulem} % sungkun added

\begin{document}

\title{Interference-enhanced optical force detection of weak light fields using a levitated nanoparticle}

\author{Seyed Khalil Alavi,\authormark{1, 2, $\dagger$,*} Youssef Ezzo,\authormark{1, 2, $\dagger$} Ashik Pulikkathara, \authormark{1, 2} and Sungkun Hong\authormark{1,2,*}}

\address{\authormark{1}Institute for Functional Matter and Quantum Technologies, Universität Stuttgart, 70569 Stuttgart, Germany\\
\authormark{2}Center for Integrated Quantum Science and Technology (IQST), University of Stuttgart, 70569 Stuttgart, Germany}

\email{\authormark{$\dagger$}These authors contributed equally to this work.}
\email{\authormark{*}Corresponding authors: skalavi@fmq.uni-stuttgart.de, sungkun.hong@fmq.uni-stuttgart.de}

\begin{abstract*} 
Optically levitated nanoparticles in vacuum provide a highly sensitive platform for probing weak light–matter interactions. In this work, we present an interference-based method to amplify the optical force exerted by a weak field on a nanoscale particle trapped in an optical tweezer. By allowing the weak field to interfere with the strong trapping beam, we significantly enhance the optical force compared to the case without interference. This amplified optical force enables the detection of the weak field through the particle's motion, reaching picowatt-level sensitivity under moderate vacuum conditions. We further discuss the potential of this approach for developing an ultrasensitive, nondestructive detector of light fields and for exploring optomechanical interactions at the single-photon level.
\end{abstract*}

\section{\label{sec:intro}Introduction}
Optical force represents a fundamental form of light-matter interaction. Arising from the momentum exchange between photons and matter, optical force has been widely explored as a fundamental and versatile tool across diverse scientific fields\cite{chu.1991.optical.trap.review}. The utilization of optical forces led to the pioneering development of optical tweezers \cite{Ashkin.1970, Ashkin.1978.tweezer}, a technique that creates a stable potential well capable of trapping and holding a wide range of physical systems, from individual atoms\cite{Chu.1998, Tannoudji.1998, Phillips.1998} to large dielectric objects including bacteria\cite{Ashkin.1987.science.virus} and single cells\cite{Ashkin.1987}. This innovation has profoundly impacted numerous scientific disciplines, from biological and medical sciences \cite{Ashkin.1987,block.1994.tweezer.bio,bustamante.2020.tweezer.protein,bustamante.2021.tweezer.biophysics} to quantum science and technology \cite{endres.2016.atom.array,barredo.2016.atom.array,bernien.2017.41.atom.array}.

This concept of optical manipulation has recently been extended to levitated optomechanics\cite{GonzalezBallestero.2021}, a vibrant field of research that explores methods for controlling the motion of mesoscopic particles levitated, for example, in an optical tweezer, in vacuum. Optical levitation in vacuum offers exceptional isolation from environmental disturbances, leading to remarkably high mechanical quality factors \cite{Chang.2010,RomeroIsart.2010}. Such high quality factors directly translate into extraordinary sensitivity to external forces and fields, establishing levitated optomechanical systems as an excellent platform for developing highly sensitive sensors \cite{Ranjit.2016, levi.sensing.PRA.2020,liang.2023.yocto.force.sense} and testing fundamental principles of quantum physics in the macroscopic domain \cite{RomeroIsart.2011}.

In levitated optomechanics, the ability to precisely tailor optical force is crucial not only for the levitation and confinement of the particle but also for actively manipulating its motion within the trap. Interference has proven to be a powerful tool in this context, offering various spatiotemporal control schemes. For instance, phase-coherent structured light has been employed to engineer scattering patterns of a large particle, thus enhancing its trapping efficiency\cite{Taylor2015_structured_light_trapping}. Similarly, counter-propagating beams have been exploited to realize an optical conveyor belt\cite{zemanek.2005.optical.conveyor.belt}, enabling long-distance particle transport through hollow-core photonic crystal fibers \cite{grass2016_fiber_transport,lindner2024_fiber_transport_uhv}. More recently, interference between the trapping beam and its reflection has been utilized to stably trap a particle in the vicinity of a surface \cite{diehl2018_levitation_near_membrane,magrini.2018,ju.2023.rotation.near.surface, alavi.2025.levi.toroid} and, combined with frequency modulation, to realize all-optical linear feedback cooling of particle's motion to its quantum ground state \cite{kamba2022_cooling_reflection}.

Despite these impressive advances, the fundamental limits of optical force generation through interference remain largely unexplored. In this work, we experimentally investigate how a minute optical field can still exert a detectable force on an optically levitated nanoparticle through interference. Specifically, we utilize the optical trapping beam as a strong coherent reference and introduce a weak signal beam to create interference between the two beams. We demonstrate that this interference significantly enhances the optical dipole force induced by the weak beam, enabling its detection via monitoring the motion of the levitated particle. Using this approach, we achieve a detection sensitivity of $37.2~\text{pW}/\text{Hz}$ at a moderate vacuum pressure of $6.8 \times 10^{-4}~\text{mbar}$. We discuss how this method could be extended to reach zeptowatt-level sensitivities, opening pathways toward nondestructive photodetection and quantum experiments on the particle's motion at the single-photon scale.

\section{\label{sec:principle}Basic Principle}
In our experiment, we consider a scenario in which a coherent weak signal beam is co-propagated with a strong tweezer field with the same frequency.
The electric fields of the two beams along the optical axis (z-axis) can be described in the following way:
\begin{equation}\label{eq:efield}
    \mathcal{E}_{tw}(z)=\hat{x}E_{tw}\cdot a(z)~e^{i\Phi(z)}, \; \mathcal{E}_{s}(z)=\hat{x}E_{s}\cdot a(z-z_{s})~e^{i\Phi(z-z_s)+i\phi_s},
\end{equation}
where $E_{tw}$ and $E_{s}$ are electric field amplitudes of the tweezer and signal beams, $z_{s}$ is the position of the focus of the signal beam relative to the tweezer beam, and $\phi_s$ is an additional phase difference between the two beams. The envelope function $a(z)$ and the phase factor $\Phi(z)$ can be modeled after the Gaussian beam:
\begin{equation}\label{eq:env}
    a(z)=\frac{w_0}{w(z)}, \; \Phi(z)= -kz +\psi(z),
\end{equation}
where $w_0$ is the beam waist at the focus, $w(z)=w_0\sqrt{1+(z/z_r)^2}$ is the beam waist variation along the z-axis, $z_r=\pi w_0^2/\lambda$ is the Rayleigh length, and $\psi(z)=atan(z/z_r)$ is the Gouy phase.

The resulting optical potential due to the two interfering beams is:
\begin{equation}\label{eq:pot}
\begin{split}
    U(z) & = -\alpha \vert \mathcal{E}_{tw}(z) + \mathcal{E}_{s}(z) \vert^2 =\alpha [E_{tw}^2a(z)^2+E_{s}^2a(z-z_s)^2 \\
    & +2E_{tw}E_{s}a(z)a(z-z_s)cos(kz_s+\phi(z))],
\end{split}
\end{equation}
where $\alpha$ is the polarizability of the particle and $\phi(z)=\phi_s+\psi(z-z_s)-\psi(z)$. The optical force $F(z)$ on the particle along the z-axis at the tweezer focus ($z=0$) in the limit of $z_s\ll w_0$ is
\begin{equation}\label{eq:for}
    F(z=0) \approx 2\alpha\frac{z_s}{z_r^2}E_s^2 + 2\alpha\frac{z_s}{z_r^2}cos(kz_s + \phi(z_s))E_{tw}E_s.
\end{equation}
Here, the first term in the equation represents the force term due to the signal field without interference $F_{s0}$, and the second term arises from the interference between the signal beam and the tweezer beam $F_{si}$. The ratio between the two force terms $F_{si}/F_{s0}\propto E_{tw}/E_{s} \gg1$ confirms the enhancement of the force induced by the interference. A more detailed derivation is provided in Section 1 of Supplement 1.

\section{\label{sec:results}Experimental Results}
\begin{figure}
\centering
\includegraphics{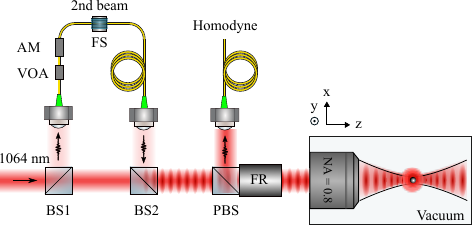}% Here is how to import EPS art
\caption{\label{fig:setup} A strong optical tweezer beam is focused by a high NA objective lens mounted inside a vacuum chamber to trap a silica nanoparticle. The particle’s scattered light is collected in the backward direction by the same objective and directed to a phase-sensitive homodyne detection system via a Faraday rotator (FR) and a polarizing beam splitter (PBS).
A secondary, weaker beam is derived from the primary tweezer beam using a beam splitter (BS1) and coupled into an optical fiber path that includes a fiber-coupled variable optical attenuator (VOA), an eletro-optic amplitude modulator (AM), and a fiber stretcher (FS). The VOA adjusts the power of the weak beam, while the AM modulates its intensity at a frequency near the particle’s mechanical oscillation frequency. The FS serves to randomize the relative phase between the tweezer and the weak beam.
The weak beam is then recombined with the primary tweezer beam using a 10:90 beam splitter (BS2). The additional optical force exerted by the weak beam induces a small perturbation in the particle’s motion, which is detected through the homodyne detection setup.
}
\end{figure}

Figure \ref{fig:setup} depicts the schematic of the experiment. Our experimental setup is centered around a standard optical tweezer setup in a vacuum chamber. The optical tweezer is formed by tightly focusing a continuous-wave 1064 nm laser using a conventional microscope objective lens with a numerical aperture (NA) of 0.8. We trap a silica nanosphere (nominal diameter of 142 nm) with the estimated tweezer beam power of $414~\text{mW}$ at the focus. The light scattered by the particle is recollected by the objective and guided to a balanced homodyne detection setup to read the motion of the particle along the z-axis. A weak laser field, derived from the same 1064 nm source, is combined with the main beam via a non-polarizing beam splitter and co-propagates with the strong tweezer field. The interference between the two fields induces an additional optical force on the particle, leading to a detectable modulation in the homodyne signal.

To controllably generate this interference-induced force, we employ several techniques. First, the intensity of the weak beam is adjusted using a voltage-controlled variable optical attenuator (VOA). To efficiently excite the particle's motion, thus to enhance the detectable signal, we modulate the weak beam's intensity near the particle’s mechanical resonance frequency using an electro-optic amplitude modulator (AM). Finally, a fiber stretcher (FS) is used to sweep and randomize the weak beam’s phase $\phi_s$, effectively averaging out the dependence of the force on the phase $\phi(z)$ shown in Eq. \ref{eq:for}. A more detailed explanation is provided in Section 2 of Supplement 1.

\begin{figure}[H]
\centering
\includegraphics{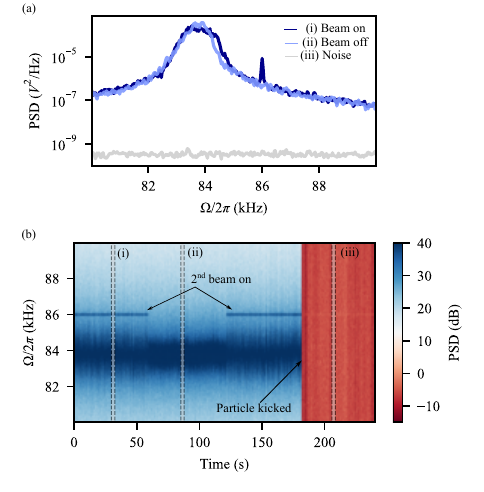}% Here is how to import EPS art
\caption{\label{fig:psd}(a) The PSDs of the detected homodyne signal at a pressure of ${0.1}~ \text{mbar}$. Traces in dark blue (i), light blue (ii), and gray (iii) are recorded when the weak signal beam is on, off, and after the particle is kicked out of the trap, respectively.  The particle's thermal motion along the z-axis appears as a broad peak around the frequency of $\Omega_z/2\pi = 83.8~\text{kHz}$. A sharp peak near the resonance is generated when the weak beam with a power of $493~\text{nW}$ is turned on and modulated with a frequency of $\Omega_{AM}/2\pi = \text{kHz}$. For each averaged PSD, time traces with the length of $\approx 33~ \text{ms}$ are first measured and converted to single-shot PSDs. A total of 35 consecutive single-shot PSDs are then averaged. (b) Evolution of the single-shot PSD as a function of time. The second beam is blocked at a time of $\sim 60~\text{s}$ and unblocked again at $\sim 120~\text{s}$. The particle is finally kicked out of the trap ($t \sim 180~\text{s}$). Horizontal markers labeled (i), (ii), and (iii) indicate the groups of the PSDs used to generate the averaged PSDs in panel (a).}
\end{figure}

Figure \ref{fig:psd}a shows averaged power spectral densities (PSD) of the detected signal measured at a pressure of $0.1~ \text{mbar}$. In the absence of the weak signal beam, a pronounced peak is observed at the frequency $\Omega_{z}/2\pi = 83.8~\text{kHz}$, which corresponds to the thermally driven motion of the paritcle along the z-axis. 
The weak signal beam, with a power of $493~\text{nW}$, is then introduced, driven by the AM with the frequency $\Omega_{AM}/2\pi = 86~\text{kHz}$. A clear and sharp peak in the PSD at $\Omega/2\pi = 86~\text{kHz}$ demonstrates an efficient transduction of the weak-beam-induced force to the particle's motion. %Here, the FS is swept with a triangular voltage signal with the frequency of $\Omega_{FS}/2\pi = 1~\text{Hz}$ and the amplitude large enough to sweep the phase more than $2\pi$. This allows us to randomize $\phi(z_s)$.

By comparing the power of the thermally driven peak with that of the peak induced by the weak beam, we first extract the root-mean-square (RMS) displacement caused by the weak-beam-induced force, $z_{si, rms} = 3.9~\text{nm}$ (see Sect. 2 of Supplement 1 for details). The corresponding RMS force exerted by the weak beam is found to be $F_{si,rms}= 145~\text{aN}$.

\begin{figure}[H]
\centering
\includegraphics{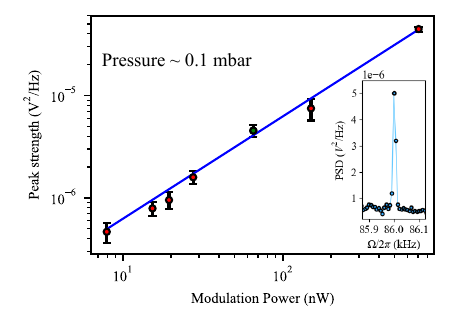}% Here is how to import EPS art
\caption{\label{fig:pwr_sweep} Signal strength of the weak beam as a function of modulation power \(P_s\), extracted from the mean peak height of the averaged PSD at the AM modulation frequency at a pressure of ${0.1}$ mbar. The inset (lower right) shows an example of the PSD around the AM frequency, which is used to obtain the data point at $P_s=66$ nW (highlighted in green). The blue line represents a fit proportional to $P_s$. The excellent agreement between the data and the fit confirms that the measured forces originate from optical interference.}
\end{figure}

Next, we sweep the amplitude of the weak beam and monitor changes in the peak strength corresponding to the driven optical force (Figure \ref{fig:pwr_sweep}). According to Eq. \ref{eq:for}, the force arising from the interference $F_{si}$ is proportional to $\sqrt{P_s}$, where $P_s$ is the power of the weak beam. We observe that the peak strength shows a clear proportionality to $P_s$. Considering that the PSD signal strength is proportional to the variance of the motion, and thus the applied force squared, the signal proportionality to $P_s$ is expected. This is clearly confirmed in Fig. \ref{fig:pwr_sweep}, validating that the measured force originates from interference. 
%. 

The high signal-to-background ratio of the peak induced by the weak beam shown in Figure \ref{fig:psd} indicates that our method can be used as a sensitive light field detection scheme. Here, we deduce the detection sensitivity of our scheme for the weak signal beam. First, from the thermal background noise at the amplitude modulation (AM) frequency, we obtain a force sensitivity of $18.5~\text{aN}/
\sqrt{\text{Hz}}$. With the calibrated $P_s=493$ nW and the estimated $F_{si,rms}= 145$ aN and considering $F_{si}\propto\sqrt{P_s}$, we conclude that our light-field detection sensitivity is estimated to be $8~\text{nW}/\text{Hz}$.
We note that our light-field detection sensitivity has a unit of $\text{W}/\text{Hz}$ instead of $\text{W}/\sqrt{\text{Hz}}$, because $P_s \propto F_{si}^2$.

\begin{figure}[H]
\centering
\includegraphics{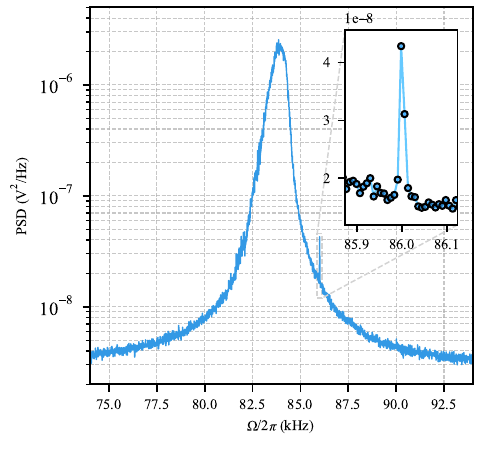}% Here is how to import EPS art
\caption{\label{fig:psd_low_pressure} The PSD of the detected homodyne signal at a pressure of $6.8 \times 10^{-4}$ mbar. The power of the weak beam is set to $58.2$ pW. A sharp peak at the AM's modulation frequency $\Omega_{IM}/2\pi = 86$ kHz clearly indicates the particle’s driven motion induced by the weak beam. From the inset, a signal-to-noise ratio of 2.6 within the measurement bandwidth can be inferred.}
\end{figure}

The force sensitivity of the particle improves with decreasing environmental pressure, as the thermal force noise due to the surrounding gas decreases. The resulting thermal-limited force sensitivity is given by\cite{levi.attonewton.PRA.2015},
\begin{equation}\label{eq:sens}
    S_{F}^{1/2}=\sqrt{4k_BTm\gamma}
\end{equation}
where $k_B$ is Boltzmann's constant, T is the environment temperature, m is the mass of the particle, and $\gamma\propto p_{gas}$ is the gas-limited damping rate, which is proportional to the gas pressure $p_{gas}$. Since $P_s \propto F_{si}^2$, we expect that the light detection sensitivity improves linearly with decreasing gas pressure. 
To confirm this, we lower the pressure of the chamber down to $6.8 \times 10^{-4}$ mbar and perform the measurement (Fig. \ref{fig:psd_low_pressure}). At this pressure level, the particle can be lost due to its increased sensitivity to other external perturbations. To avoid this, we stabilize the particle's motion by employing linear electrical feedback cooling\cite{levi.electrical.cooling.PRL.2019}. 
We observe that the weak beam of the power down to $58.2$ pW can be measured with the signal-to-noise ratio of 2.6 compared to the noise floor.
From this, we deduce the light detection sensitivity of $37.2~\mathrm{pW}/\text{Hz}$. It shows a significant improvement in the sensitivity compared to that of $8~\text{nW}/\text{Hz}$ obtained at $0.1$ mbar. 
The sensitivity enhancement factor of 215 agrees well with the corresponding pressure ratio of 141, considering that the accuracy of the pressure gauge used in the experiment (WRG-S - NW25, Edwards Vacuum) is approximately 30 \%.

\section{\label{sec:conclusion}Conclusion}
In summary, we demonstrate that in a levitated optomechanical system, the optical force exerted by a weak light field can be significantly enhanced through its interference with the strong optical tweezer field. We validate the nature of interference-enhanced force by observing the linear dependence of the measured noise power of the force on the power of the inserted weak light beam. Moreover, this interference-induced enhancement of the force, combined with the excellent force sensitivity of a vacuum-levitated nanoparticle, enables the detection of the weak field with the sensitivity of $37.2~\mathrm{pW}/\text{Hz}$ under a pressure of $6.8 \times 10^{-4}$ mbar.

\begin{figure}[H]
\centering
\includegraphics{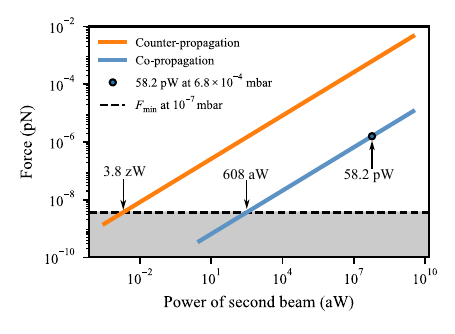}% Here is how to import EPS art
\caption{Expected interference-induced force as a function of power of the second beam, when the second beam is co- (blue) or counter- (orange) propagated with respect to the primary tweezer beam. The force in the co-propagation mode is calculated using the same experimental parameters demonstrated in this work. The force in the counter-propagation mode is increased by two orders of magnitude compared to the co-propagation mode for the same level of the second beam power. The gray area and the dashed line show the minimum detectable force, $F_{min} = \sqrt{\tau S_F}$ at a pressure of $10^{-7}~\mathrm{mbar}$ for a measurement time, $\tau = 1$ s. At this pressure, detection of a beam with a power of down to $3.8$ zW is feasible with the counter-propagation configuration.    
\label{fig:psd_low_pressure_theory} }
\end{figure}

We expect that further improvements to our experimental setup and configuration will enable us to explore several interesting directions. 
First, as already shown in Fig. \ref{fig:psd_low_pressure}, reducing the pressure can improve the sensitivity of our scheme further. For instance, at a pressure of $10^{-7}$ mbar, where the environmental force noise is still largely limited by the gas collision\cite{levi.recoil.PRL.2016}, the light field sensitivity of the current setup is expected to reach $608~\text{aW}/\text{Hz}$ (Fig. \ref{fig:psd_low_pressure_theory}).

Moreover, reversing the propagation direction of the weak beam to counter-propagation can further enhance the interference-induced optical force by two orders of magnitude (see Sect. 1 of Supplement 1 for details).
With this improvement, we anticipate that the light field as weak as $3.8~\mathrm{zW}$ can be measured with a bandwidth of 1 Hz (Fig. \ref{fig:psd_low_pressure_theory}). $3.8~\mathrm{zW}$ corresponds to the mean photon flux of about 0.02 photons per second with a wavelength of 1064 nm. We also emphasize that our approach does not involve the annihilation of the light field. This feature opens the possibility of developing a nondestructive photodetection scheme\cite{reiserer.2013.photodetec.nondestructive,niemietz.2021.photodetect.nondestructive,distante.2021.photodetect.nondestructive} based on a levitated sensor, with its sensitivity approaching the single-photon level. 

Integrating a high-finesse optical cavity could further enhance the interaction strength. For instance, employing a cavity with a finesse of $10^5$ would enable the detection of a single photon within a measurement time of $10~\mu s$, which is comparable to the oscillation period of a typical levitated nanomechanical resonator. This capability would offer a unique platform for exploring optomechanical interactions at the single-photon level.

\begin{backmatter}
\bmsection{Funding}
Deutsche Forschungsgemeinschaft (Projektnummers: 523178467). Carl-Zeiss-Stiftung (Johannes-Kepler Grant through IQST).

\bmsection{Disclosures}
The authors declare no conflicts of interest.

\bmsection{Data Availability}
 The data that support this study are available upon reasonable request from the authors.

\bmsection{Supplemental document}
See Supplement 1 for supporting content. 

\end{backmatter}

\bibliography{sample}

%%%%%%%%%% If preparing manually:
% \begin{thebibliography}{1}
% \newcommand{\enquote}[1]{``#1''}

% \end{thebibliography}

\end{document}